\documentclass[12pt]{article}
\usepackage{amsfonts}

\newcommand\cA{\mathcal{A}}
\newcommand\cO{\mathcal{O}}
\newcommand\cP{\mathcal{P}}
\newcommand\fA{\mathfrak{A}}
\newcommand\al{\alpha}
\newcommand\om{\omega}
\newcommand\De{\Delta}
\newcommand\beq{\begin{equation}}
\newcommand\eeq{\end{equation}}

\begin{document}
\title{Questions in quantum physics: a personal view}
\author{Rudolf Haag\\[1mm]
Waldschmidtstra\ss e 4b, D--83727 Schliersee-Neuhaus, Germany}
\date{}
\maketitle

\begin{abstract}{
An assessment of the present status of the theory, some immediate 
tasks which are suggested thereby and some questions whose answers may
require a longer breath since they relate to significant changes in the conceptual  and mathematical structure of the theory. }
\end{abstract}
\section{Introduction}
Personal views are shaped by past experiences and so it may be worth while 
pondering a little about accidental circumstances which channelled the course of 
one's own thinking. Meeting the right person at the right time, stumbling across
a book or article which suddenly opens a new window.

Fifty years ago, as eager students at Munich University just entering the phase
of our own scientific research, we were studying the enormous papers by Julian
Schwinger  on Quantum Electrodynamics, following the arguments line by line but not
really  grasping the message. I remember the feelings of frustration, realizing that we
were far away from the centers of action. But, mixed with this, also some dismay 
which did not only refer to the enormous arsenal of formalism in the new developments 
of QED but began with the standard presentation of the interpretation of Quantum 
Theory. I remember long discussions with my thesis advisor Fritz Bopp, often
while circling some blocks of streets ten times late in the evening, where we
looked in vain for some reality behind the enigma of wave-particle dualism.
Why should physical quantities correspond to operators in Hilbert space? Why 
should probabilities be described as absolute squares of amplitudes?, etc., etc.
Since we did not seem to make much headway by such efforts I decided
to postpone philosophy and concentrate on learning what was really done, 
which aspects were used in an essential way and were responsible for the
miraculous success of Quantum Theory. Leave aside for a while questions of
interpretation discussed by Bohr and Heisenberg, extrapolations like Dirac's
transformation theory or von Neumann's theory of measurement and return to
the more pragmatic attitude pervading for instance the book by Leonard Schiff
on Quantum Mechanics. For this purpose WaIter Heitler's book `\emph{Quantum Theory
of Radiation}' (second part of the first edition) proved immensely helpful. Here
I saw what the typical problems were which Quantum Field Theory tried to address 
and I also learned to appreciate the progress made meanwhile by covariant 
perturbation theory and Feynman diagrams.

The next great piece of luck for me was (indirectly) caused by Niels Bohr.
In the course of the planning of a great joint European laboratory for high energy 
physics (now CERN) he saw the need to introduce a young generation of theoretical 
physicists in Europe to this area and offered the hospitality of his Institute.
He called an international conference in 1952 which I could attend and a year 
later I got a fellowship for spending a year in Copenhagen. This clearly ended
the frustration of being isolated from the great world. The first fringe benefit
of the 1952 conference was a garden party at the residence of Niels Bohr where
I met Arthur Wightman who gave me the invaluable advice to read the 1939 
paper by Wigner on the representations of the inhomogeneous Lorentz group. 
Returning to Munich I followed the advice and it was a revelation. Not only
by putting an end to our concern about wave equations for particles with higher 
spin but because here I recognized a most natural starting point. The group, 
nowadays called the Poincar\'e group, is the symmetry group of the
geometry of space-time according to the theory of special relativity. What
would be more natural than to ask for the irreducible representations of this 
group? Equally remarkable was the result that these representations (more precisely 
those of positive energy) correspond to the quantum theory of the simplest physical 
system, a single particle. It has just two attributes: a mass and a spin. Everything 
that can be observed for such a system, including, as far as possible, a position
at given time, could be expressed within the group algebra. No reference to 
canonical commutation relations, guessed from classical mechanics, nor to
the wave picture. The wave equation arises from the irreducibility requirement.
I could not understand why this paper had remained almost unnoticed by the 
physics community for such a long time. In fact, even in 1955 I was introduced
at some conference in Paris as: `He is one who has read the 1939 paper
of Wigner'. This was indeed my major claim to fame.

Probably all the young theorists who had the privilege of spending a year
as members of the CERN theoretical study group in Copenhagen will remember 
this as a wonderful time. The atmosphere at the Institute with the \emph{spiritus
loci}, emanating from the personality of Niels Bohr, and upheld by Aage Bohr and
Christian M\o ller who was the official director of the study group, combined
in a rare way scientific aspirations on the highest level with a friendliness
discouraging any competitive struggles and allowing everyone to proceed at his 
own pace. Though there was some joint topic suggested, at my time it was the
Tamm-Dancoff method and alternatives to perturbation theory, with some emphasis 
on nuclear physics, everyone was allowed to work on the subject of his own 
choice. So, after a short excursion into nuclear physics I returned to my pet 
subjects. Prominent among them was collision theory in quantum mechanics
and field theory. The widely used recipe of `adiabatic switching off of the
interaction' appeared to me not only as ugly but also as highly suspect because the
notion of `interaction' was not clear \emph{a priori} and if one switched off the wrong
thing one would decompose a nucleus into its fragments. This led me to appreciate
the physical significance of various topologies for vectors and operators in Hilbert 
space. Since in high energy physics the experiments were not concerned with 
fields but with particles there was the idea that the role of a field was just to 
interpolate between incoming and outgoing free fields which were associated to 
some specific type of particle. Trying to implement this I unfortunately used
the wrong topology. But just at the end of my stay in Copenhagen we received 
a preprint of the paper by Lehmann, Symanzik and Zimmermann who did things 
correctly and thereby provided an elegant algorithm relating `Green's functions'
in quantum field theory to scattering amplitudes for particles.

Speaking of Copenhagen and quantum theory the `Copenhagen interpretation' 
immediately comes to mind. I prefer to call it the `Copenhagen spirit' or, more 
specifically, the natural philosophy of Niels Bohr. I did have some opportunities
to talk to the great master but, in spite of my great admiration and some efforts, 
this was not fruitful. It was only in later years that I understood the depth
of various parts of his philosophy. But there always remained one disagreement 
which came from the question: what are we trying to do and what is guiding
us? Physics began by the recognition that there are relations between
phenomena which are reproducible. These could be studied systematically,
isolating simple processes, controlling and refining the conditions under which 
they occur. The formulation of the regularities found and the unification of
the results of many different experiments by one coherent picture was achieved
by a mapping into abstract worlds: a world of appropriate concepts and a world
of mathematical structures dealing with relations within and between various sets 
of mental constructs, one of them being the set of complex numbers. This endeavor 
manifestly led to some level of understanding of `the laws of nature' as evidenced 
by the development of a technology which provided mankind with enormous powers 
to serve their conveniences and vices. But what was understood and what was
the relative role in this process played by observations, by creation of concepts 
and by mathematics? When Dirac wrote on the first pages of the 1930 edition of his 
famous book: `\emph{The only object of theoretical physics is to calculate results that 
can be compared with experiment}', this can hardly be taken at face value. As
he often testified later, he was searching for beauty and he found it in
mathematical structures. So much so that he preferred to look for beautiful 
mathematics first and consider their possible physical relevance later.
Indeed, the road from phenomena to concepts and mathematics is not a one-way 
street. As the studies shifted from coarser to finer features the theory could
not be derived directly from experiments but, as Einstein put it, it had to be freely 
invented and tested subsequently by suggesting experiments. In this passage
back and forth between phenomena and mental structures many aspects entered 
which cannot be rationalized. The belief in harmony, simplicity, beauty are driving 
forces and they relate more to musicality than to logic or observations.

There was one further highly significant and somewhat accidental occurrence
in shaping my subsequent work and this may also illustrate the above remarks. 
Professors F. Bopp and W. Maak in Munich decided in 1955 that it was
important to exchange experience between theoretical physicists arid
mathematicians. This initiative was not rewarded by visible success. The number
of participants dwindled quickly and the enterprise ended after a few months.
But for me it was of paramount importance. I was introduced there to rather recent 
work of the Russian mathematicians Gelfand and Naimark on involutive, normed 
algebras and to the work of von Neumann on operator algebras and reduction 
theory. I saw that Hilbert space resides in some wider setting which, at least 
from the mathematical point of view, constitutes a rather canonical structure 
resulting from a few natural structural relations. Besides the standard algebraic 
operations it needed a *-operation (involution) and one is led to a natural
topology induced by a unique `minimal regular norm'. It appeared highly likely 
that this structure was behind the scenes in the mathematical formalism of 
quantum theory. The prototype of such an algebra is furnished by group theory.
There a most important tool is the consideration of functions of the group
with values in the complex numbers. They form obviously a linear space because 
we can multiply them by complex numbers and add them. If there is a
distinguished measure on the group (which is the case for compact groups
and, up to a normalization factor, for locally compact ones) the group
multiplication defines a product in the space of these functions, the convolution
product. The inverse in the group defines a *-operation in this algebra of functions 
by $f^*(g) = \bar f(g^{-1})$. The resulting algebra yields the representation theory of
the  group. An irreducible representation corresponds to a minimal (left) ideal in the 
algebra.

If in the following I try to describe things which I believe to have learned 
concerning the interrelation between observed phenomena, concepts and mathematical 
structures I must precede this with some apologies. The inadequate handling of 
references is due to the state of disorder in my notes and lack of time.
The abstractions used in describing the procedures of acquiring knowledge may 
be too schematic. There is a painful gap between their qualitative character and 
the very precise mathematical structures into which they are mapped.

\section{Concepts and Mathematics in Quantum Theory}
A paper describing some fascinating recent experiments was entitled
`\emph{Reality or Illusion?}' These experiments (see e.g.\ refs.
\cite{ref1,ref2,ref3})
have lent impetus to the long standing discussion about the meaning of reality in
quantum theory. Do  the discoveries force us to abandon the naive idea of an outside
world called  nature whose laws we try to find? What is the role of the observer? Do the
puzzles  relate to the mind-body problem?

Many different views concerning such questions have been voiced throughout
the past seventy years. So if I try to express mine it may be pardonable to proceed 
in an extremely pedantic fashion.

A single experiment of the type alluded to above combines many individual 
clicks of some detectors. Though each click is unique and neither repeatable
nor predictable even under optimally controlled circumstances, we may regard
it as a `real fact' in the sense in which these words are used in any other context. 
Let us call it an `event'. Its existence is not dependent on the state of consciousness 
of human individuals. In modern times it is usually registered automatically,
stored in computer memories and there is no dispute between the members of
a group of experimenters about its `reality'. The outcome of the experiment refers
to the frequency with which a particular configuration of events occurs in many 
runs and this is reported as a probability of the phenomenon under precisely 
described circumstances. To be accepted, this result must be reproducible by any 
other group of scientists who is willing to invest time and resources in repeating the 
experiment. The events mentioned are coarse. A detector is macroscopic.
We regard macroscopic bodies as `real objects' and statements about their placement 
in space and time as `real attributes'. The word `real' just means here that
such objects, events and their space-time attributes belong to common experience 
shared by many persons and do not depend on the state of consciousness of
an individual. The observed relations between them constitute the only empirical 
basis from which the `free invention' of a theory can proceed. With regard to this 
mental task there is a piece of wisdom which I learned from F. Hund. It might
be called Hund's zeroth rule. He pointed out that the progress of physical theory 
depended on the lucky circumstance that always some effects were small enough to
remain unnoticed or could be disregarded as insignificant at the time a particular 
piece of the theory was proposed. We cannot take many steps at the same time.
We should regard a theory always as preliminary; it will disregard some fine 
features of which we are luckily ignorant or which we neglect in order to obtain
a tractable idealization.

The purpose of these lengthy elaborations is twofold. First, I do not think 
that physics can make any contribution to the mind-body problem. The attempt
to explain some puzzling aspects of quantum physics by invoking subjective
impressions and the role of the consciousness of individual human beings
is not an appropriate answer. Secondly, the concept of \emph{event} is necessary in 
quantum physics. It is an independent concept. The mental picture that it corresponds 
to an interaction process between an atomic object and a macroscopic one is 
misleading because experiments tell us that there are no atomic objects in an 
ontological sense (see below). Of course, from this we may conclude that there 
are no macroscopic objects either and that their apparent reality results from an 
asymptotic idealization. This is true (see e.g.\ the discussion in
ref. \cite{ref4} of
the emergence of classical concepts due to large size and decoherence). But the
idealization is covered by Hund's zeroth rule and is essential for the form of the
present theory. If we want to avoid it we must take the next step in the development
of the theory. Let us address now some specific aspects.

1. In experiments we usually (necessarily?) distinguish two parts: a source which 
determines the probability assignments (subsumed under the notion of `state'), 
and a set of detectors to whose responses the probabilities apply. Though in
the description of the source a variety of considerations enter which will have
to be looked at more closely, we shall, for simplicity of language, just idealize
it as characterized by some pattern of `source events'. The total setting, consisting 
of source events and target events, where the former determine a conspicuous 
probability assignment for the occurrence of the latter, may be called a
`quantum process'. Bohr emphasized the indivisibility of the process as
one of the key lessons of quantum theory. This poses the question of how we
can isolate such a process from the rest of the world. In technical terms, what
do we have to take into account in `preparing a state' in order to get a reproducible
probability assignment for a pattern of target events (defined by some arrangement 
of detectors). Here we are helped by lucky circumstances. We live in a reasonably 
steady environment; its influence does not change rapidly in space and time.
So, if we are stupid in the state preparation we just get an uninteresting probability 
assignment, a `very impure state'. The art of the experimenter is needed
to improve state preparation and render the probability assignment as conspicuous 
as possible. It appears, however, that there is a limit to such improvements, idealized 
by the notion of a `pure state'.

2. Given some definite process one would like to assign to it a `physical system'
as the agent producing the target events or, more carefully, as the messenger 
between source and target events. This is clearly a mental construct. Can we 
attach any element of reality to it? If we focus on a single event involving one 
detector far removed from the source we may think of a single particle as
this messenger. But we may also consider patterns of several events, seen in 
coincidence arrangements of detectors far removed from the source and from 
each other. Then we sometimes find correlations in their joint probabilities which 
are of a very peculiar type. If we believe that there is a specific messenger from 
the source to each target event (for instance a particle) then, whatever notion of 
state we try to assign to those, we cannot represent the joint probability for
the pattern of events as arising from joint probabilities for a corresponding
set of individual states of the messengers. This is the conclusion to be drawn 
from the violation of Bell's inequality. It is not so easily seen in the first
discussions which focused on hidden variables but emerges clearly in
ref. \cite{ref5}.
Another equally surprising effect has been demonstrated by Hanbury-Brown
and Twiss. They start from two entirely independent source events (for instance 
photons emitted from two far distant surface regions of a star which happen
to arrive almost simultaneously in the observatory). So they can cause a
coincidence in two detectors. Each detector responds to one photon but can, of 
course, not distinguish from which source event it comes. By varying the difference 
of the optical paths from the telescope exit to the two detectors one finds varying 
intensity correlations in the coincidence signals. This means that the cause for
the response of one detector cannot be attributed to the arrival of either a
messenger from the left edge or from the right edge of the star. Both photons 
work together though there is no phase relationship between the two emission acts. 
There is a causal relation between the pattern of two source events and the pattern 
of two target events but it cannot be split into causal ties between single events. 

Taken together these experiences imply that the notion of a `physical system' 
does not have independent reality. What is relevant for the click of a single 
detector is some notion of `partial state' prevailing in its neighborhood. In both 
of the above examples this is described as an impure state of a single particle.
In the EPR-example as discussed by Bell it is determined by one source event,
the decay of an unstable particle. In the second example it is caused by two source 
events. The probability for the response of both detectors in coincidence depends 
on the partial state in the union of the two neighborhoods and this is apparently 
not determined by the pair of partial states around the individual detectors.
Thus quantum physics exemplifies the saying: `the whole is more than the sum of 
its parts' and it does so in extreme fashion. The Pauli principle claims that
all electrons in the universe are correlated. The reality behind the mental picture 
of a physical system consisting of a certain number of particles refers to a
certain set of events with causal connections between them, manifested
by the existence of a probability for the total process. In an ideal experiment 
this is obtained by counting the number of times the pattern of target events is 
realized, dividing it by the number of times the source events occured.
In the usually prevailing cases where the source is not adequately
known we can still determine relative probabilities of different patterns of 
target events assuming that the source remains constant.

The holistic aspect mentioned above is often called the `essential non-locality'
of quantum theory.  But this is an unfortunate terminology because the only
reason why we can talk about specific processes at all resides in the locality of 
individual events and the causal structure of space-time.

3. The reader may have wondered why I specialized the usual notion of `observable' 
to that of a detector and talked about events instead of measuring results indicated 
by the position of a pointer of some instrument. The spectral resolution of
self-adjoint operators which played such an essential role in the development
of early quantum mechanics was not even mentioned so far. One reason for this 
is the problem of how to achieve the mapping from a particular arrangement of 
instruments to its representative in the mathematical scheme. In early quantum 
mechanics the idea that we consider a physical system consisting of a definite number 
of particles seemed to pose no problem (a beautiful illustration of
Hund's zeroth  
rule). The degrees of freedom were positions and momenta, appearing in the
canonical formalism of classical mechanics on equal footing. Though it became clear 
that these degrees of freedom could not be real attributes of the system one
still talked about measuring one of them (or simple functions of them like energy and 
angular momentum). Bohr emphasized that the full description of the experimental 
arrangement is needed `to tell our friends what we learned' and that this could
only be done in plain language. But, since the classical degrees of freedom persisted, 
this description of the arrangement could ultimately be summarized by one mathematical 
object which corresponded in a symbolic way to a classical quantity. How can
one proceed in this passage from the description of an arrangement of hardware
to a mathematical symbol relating to the system? The primary piece of information
is given by the placement of macroscopic bodies in space-time. These bodies
perform different functions. Some parts may be considered as `state preparation 
procedures' representing the source events. Other parts yield the measuring
result which is an unresolvable phenomenon, an \mbox{unpredictable}  
\mbox{decision} in nature,
a coarse event. Its primary attribute is an approximate position in space-time.
The representation of the whole arrangement (apart from the primary source)
by a self-adjoint operator, interpreted as describing the measurement of a
quantity related to some function of the classical degrees of freedom involves
the theory (Schr\"odinger equation) in conjunction with idealizations and
approximations which are transparent only in simple cases. The operators corresponding
to momentum
and energy have clear significance as generators of translations in space and
time but are only indirectly related to observations, which in the last resort
concern the position of an event in space-time. The position operator of a
particle at a prescribed time yields spectral projectors which can approximately 
characterize an event. But the assumed existence of a family of mutually exclusive 
events with certainty that one of them must happen is an extrapolation which
becomes highly unnatural in relativistic situations. This is mildly indicated already
by the ambiguities arising in the attempt to define a position operator in Wigner's 
analysis.

The fundamental discovery, that the `elementary particles', formerly
believed to be the building stones of matter, are not eternal but can be created 
or destroyed in processes, forces us to consider states whose particle content
is not only varying but undefined in some regions. While the concepts of `system' 
or `particle' suggest some object existing in an ontological sense, the concept of
`state' belongs to the realm of possibilities (potentialities, propensities) for the
realization of \emph{something coming into existence}, an event. This would not be
so in a deterministic theory but if we believe that the indeterminacy in the prediction 
of phenomena, inherent in the formulation of quantum physics, is a feature of
the laws of nature and not just due to ignorance which could be lifted by future 
studies then the distinction between the realm of possibilities and the realm
of facts becomes imperative. The `state' belongs to the former. Strictly speaking
it provides a quantitative description of a \emph{contribution} to the probability for
the occurrence of events. The other contribution is given by the placement and 
type of detectors. Thus also the notions of `system' and `particle' belong to
the realm of possibilities. But they retain their importance. They allow us
to classify (at least under favorable circumstances) the possible partial
states in a region by a \emph{denumerable set}.

This procedure involves the center piece of the mathematics of quantum theory:
the superposition principle and eigenvalue problems; more generally, the determination 
of invariant subspaces in a complex linear space with respect to the action of the 
symmetry group of the theory. The intuitive steps leading to the recognition that 
this mathematical structure (Hilbert space, involutive algebras, representation theory 
of groups) offers the key to quantum theory appear to me as a striking
corroboration of Einstein's emphasis on free creations of the mind and Dirac's 
conviction that beauty and simplicity provide guidance.

Returning to our line of argument: the ordering of states in classes by the
concept of a `system' corresponds to the selection of an invariant subspace, 
under the action of the symmetry group. A minimal invariant subspace, an
irreducible representation, may be called an elementary system. Its attributes are 
group characters. If we consider only the symmetry group of space-time,
the Poincar\'e group, the irreducible representations give us states of a
stable system, a system which could persist eternally if it were alone in
the world and no events could occur. This simple system is a single particle.
Its attributes are a value of the mass and the spin, which define a group
character. The reason why the simplest systems play such an important role
for observations is due to the circumstance that in many experiments the
partial state pertaining to a large but limited region of space-time can
be very closely approximated by the restriction of \emph{a global single particle
state} to the region. This will, in fact be the standard situation in the overwhelming 
part of space-time if the mean density of matter is small. To obtain a basis in the 
space of single particle states we must choose some maximal set of commuting 
generators. In this choice the generators of space-time \emph{translations}, whose
spectral  values are energy-momentum 4-vectors, play a preferred role in the following 
respect. If we look in a region whose extension is small compared to its mean 
distance from source events then the partial state there is well approximated
by a mixture of parts of plane waves, (improper) eigenstates of the
generators of translations, each  
belonging to a specific energy-momentum vector.

We confined attention so far to the Poincar\'e group describing the space-time 
symmetry. The full symmetry group includes `gauge symmetries' whose characters 
are charge quantum numbers. The first example was the electric charge with
its description as a character of U(1). The generalizations of this in high energy 
physics led to flavor and color multiplets associated with the groups SU(2),
SU(3). To avoid misunderstandings it must be stressed that we talk here about
a global gauge group. The significance of local gauge invariance will be addressed 
later.

\subsection{Conclusions}
Position and momentum belong to different parts of the scheme. Position is
an (approximate) attribute of an event, not of a particle, and the event marks 
a position in space-time not a position in space at an arbitrarily assumed time
(as the picture of a world line for a particle would suggest). In simple cases the 
event may be regarded as the interaction process between a particle and a detector.
But the notion of `particle' does not correspond to that of an object existing in any 
ontological sense. It relates to the simplest type of global state and describes 
possibilities, not facts. The notion of `partial state' demands in addition
that we ignore all possible events outside some chosen region and thus ignore 
possible correlations with outside events. The concepts of `particle' and
`physical system' arise from the possibility of ordering global states into
distinct classes defined by the symmetry group of the theory. A particle
corresponds to an irreducible representation of this group. Its attributes are
group characters. A system corresponds to some subrepresentation of the tensor 
product of irreducible representations. Experience tells us that only a countable
set of irreducible representations (particle types) appears in nature. The determination 
of these (the masses, spins, charge quantum numbers of physical particles) is one 
of the tasks of the theory.

In observations we are concerned with partial states which result by the
restriction of global states to some regions in which we choose to place
detectors. For a fixed global state the partial state in a region can be approximately 
described by the restriction of a global state which belongs to the class of some 
specific system. In other words: if we focus attention on some particular region then 
the global state may tell us for instance that in there the probability for events is 
almost the same as that predicted from the restriction of some single particle 
state. Indeed, if we choose the region sufficiently small then it will usually suffice 
to consider only mixtures of single particle states with definite momenta. The 
existence of zero mass particles complicates this picture somewhat, as evidenced
by laser beams and by infrared problems where the number of particles is no 
longer useful for the description of a partial state.

The analysis of global states in terms of various systems approximating the
partial states in various regions of space and time is the other task of the theory 
(the theory of collision processes).

In the whole scheme we still need an observer. No facts are created if no 
detectors are around anywhere. Though the consciousness of an individual plays no 
role (it was eliminated by the assumed `as if' reality of macroscopic bodies 
and coarse events) the scheme still appears somewhat artificial. It is a description
of what we may learn by experiments. But looking at the detailed mathematical
structure, developed to cope with the above mentioned tasks of the theory,
it seems clear that the notions of macroscopic bodies and coarse events are asymptotic 
concepts. If, on the other hand, we wish to replace them by finer ones we encounter 
difficulties. They can, I believe, not be overcome without a radical change of the 
formalism involving our understanding of space and time. As long as there is an 
enormous disparity between collision partners, one being a macroscopic body the other 
an atomic object, we can talk about an approximate position of the event and give 
upper bounds for its uncertainty, relating to the size and the time of sensitivity of 
the (effective part of the) detector, and we can give lower bounds for the energy-
momentum transfer needed to overcome the barriers against the appearance
of a significant change. This suffices for practical purposes but does not seem
to be the ultimate answer if we look at the vertex of a high energy event in a 
storage ring.

\section{The mathematical structure in relativistic quantum physics and its
interpretation}
In quantum field theory the basic mathematical objects, the fields, are functions 
of points in space-time. These are singular objects which have to be smeared 
out over some finite regions to yield observables which can be represented
by operators in a Hilbert space. There are problems. Some serious ones are related 
to gauge invariance, specifically to the local gauge principle first encountered
in quantum electrodynamics (QED). If one wants to avoid `unphysical states'
one has to restrict attention to gauge invariant quantities. From these one may 
hope to construct algebras of observables. To be precise: we abstract from this
heuristic consideration that we can obtain a normed, involutive algebra (for short,
a $C^*$-algebra) for each bounded, open region $\cO$ of space-time. The correspondence
 \beq
\cO\to\cA(\cO)
\label{eq1}
 \eeq
between regions and algebras yields one essential piece of information for the 
analysis of the consequences of the theory. We call $\cA(\cO)$ the algebra of
observables  of the region $\cO$. There are some natural relations between these local
algebras.  Obvious is the inclusion relation:

($i$) $\cO_1\subset\cO_2$ implies $\cA(\cO_1)\subset\cA(\cO_2)$.

\noindent
This allows the definition of a global $C^*$-algebra $\fA$ as the `inductive limit',
the completion of the union of all local algebras in the norm topology.

The second important relation reflects the causal structure of space-time: 

($ii$) If $\cO_1$ is space-like to $\cO_2$ then $\cA(\cO_1)$ and $\cA(\cO_2)$ commute.

The third basic ingredient is covariance with respect to the Poincar\'e group $\cP$.
We need a realization of $\cP$ by automorphisms of $\fA$; to each element $g\in\cP$
there  is an automorphism of $\fA$ denoted by $\al_g$ which should have the obvious
geometric significance:

($iii$) If $A\in\cA(\cO)$, then $\al_g A\in\cA(g\cO)$,

\noindent
where $g\cO$ denotes the region resulting from shifting $\cO$ by $g$.
It is convenient to assume that these algebras have a common unit element.

We call the structure defined by the correspondence (\ref{eq1}) with the properties 
mentioned a (covariant) net of local algebras. A general \emph{state} $\om$ corresponds
to  a normalized, positive linear form i.e.\ a linear function $A\to\om(A)$ from the
algebra $\fA$ to the complex numbers which takes real, non-negative values on the
positive elements of the algebra:
 \beq
\om(A^*A)\ge 0\;\mbox{for any}\;A\in\fA;\qquad\om(\mathbf{1}) = 1.
\label{eq2}
 \eeq
A partial state in some region $\cO$ is defined in the same way with $\fA$ replaced
by $\cA(\cO)$. It corresponds to the restriction of a class of global states to the 
subalgebra considered.

From section 2 we see that the physical interpretation requires a characterization 
of those elements of $\fA$ which represent detectors for an event in a region $\cO$.
The first guess might be to identify the projectors in $\cA(\cO)$ with such detectors.
This is, however, not sufficient. A detector, in contrast to a source, must be
passive; it should not click in the vacuum situation. We must control the energy-
momentum transfer. Starting from any element$A\in\fA$ we can construct elements
 \beq
A(f) = \int(\al_x A) f(x)\, d^4x,
\label{eq3}
 \eeq
where $x$ refers to a translation in space-time. If the Fourier transform of the 
function $f$ has support in a region $\De$ in $p$-space then the energy-momentum
transfer  of $A(f)$ is limited to $\De$. Therefore we add to the structure described so
far the (somewhat over-idealized) assumption that there exists a ground state $\om_0$,
the vacuum, which is invariant with respect to the Poincar\'e group and is 
annihilated by any $L\in\fA$ which is of the form (\ref{eq3}) with support $\De$ outside
of the closed forward cone $V^+$ (positive time-like vectors in $p$-space including
$0$). This assumption, called the \emph{spectrum condition}, allows us to define
detectors which are approximately associated to a region $\cO$ in position space and
to some window $\De$ in $p$-space indicating the minimal energy-momentum of the
`atomic object' needed for the response of the detector. Any element
 \beq
P=L^*L\quad \mbox{with}\quad \| P\| =1,\quad L= A(f)
\label{eq4}
 \eeq
represents such a detector if we start with $A\in\cA(\cO)$ and choose the function $f$ 
so that (apart from small negligible tails) the support in $x$-space is a small 
region around the origin and its Fourier transform is practically zero outside
of the region $\De$. Of course this does not yield a precise localization or momentum 
transfer but this is not relevant if we think of a detector as a macroscopic 
body. Let us note that $P$ will in general not be a projector but this is not 
necessary either, because we need not consider the negation, an instrument which 
indicates with certainty that no event has happened.

The above characterization of a detector does not tell us what the detector 
detects. But we discover that such additional information is not needed \emph{a priori}.
If the net is given and a vacuum state exists (the spectral condition), then
we have the tools to analyze the physical content of the theory by studying
the response of coincidence arrangements, represented by products of $P_k$ belonging 
to mutually space-like situated localization regions, in any state. A single particle 
state, for instance, can be defined as a state which is `simply localized at all 
times', i.e.\ never capable of producing a coincidence of two (space-like separated)
detectors:
 \beq
\om(P_1P_2)=0
 \eeq
for any such choice of the $P_k$ ($k =1, 2$), but $\om(P)\neq 0$ for some $P$. For
further elaborations see \cite{ref6}.

A net satisfying only the requirements mentioned so far need not yield a physically 
reasonable theory. It may, for instance, describe no particles at all or a
non-denumerable number of different types. Further properties are needed. Some 
necessary conditions are known which concretize the structure considerably and  relate
to various physical aspects ranging from the appearance of charge quantum  numbers in
particle physics to properties of thermodynamic equilibrium states \cite{ref6,ref7}. 
But we do not know yet how to formulate restrictive conditions powerful enough to
define a specific net, let alone the ambitious aim of constructing a net whose 
physical content is corroborated by experiments.

It is my personal conviction that in this step the local gauge principle plays
a crucial role. This assessment stems partly from progress in theoretical
high energy physics in the past decades and partly from my belief in simplicity
and naturalness of fruitful basic concepts. The principle mentioned tells us that
we should not try to focus on global symmetries. In a local theory the symmetries 
should only govern the structure in the small and the comparison of their action
in different regions needs additional information which is called a `connection' 
because the comparison depends on the way we pass from one region to the other.
In the two important classical field theories which have proved their worth for 
physics, Maxwell's electrodynamics and Einstein's general relativity, this
principle is encoded. In the former it was recognized rather late and refers to the 
gauge symmetry related to electric charge; in the latter it was one of the guiding 
principles and refers to the Poincar\'e symmetry of space-time. The Lorentz part, 
which keeps one point in space-time fixed, is reduced to a local symmetry for
the tangent space at this point; the translations are replaced by the connection. 
Quantum physics as we know and use it is anchored on the uncritical acceptance
of space-time as an arena in which we can place instruments, an arena with
known geometry including a causal structure. Some aspects of the theory depend 
also on the existence of a global symmetry for this geometry.\footnote{Quantum physics
in `curved space-time' (representing a given, external  gravitational field) retains
the first part of these requirements. This means that  the net structure of local
algebras with the properties ($i$), ($ii$) persists. The loss  of ($iii$) implies that
the spectrum condition has to be replaced. A considerable  amount of work has been
devoted to this problem but we shall not discuss it here; it would be beyond the scope
of this paper.}   If we loose this anchor completely we enter an area in which the
conceptual structure and  mathematical formalism of quantum physics cannot persist. In
this area the  problem mentioned at the end of section 2 and some of the questions
addressed in the next section may become imperative. So, to stay on the present level,
we wish to keep global Poincar\'e symmetry and only reduce the internal symmetries, 
relating to the charge structure, to local significance needing the definition of
a connection. In a classical field theory the formalism of Yang-Mills theories, 
generalizing electrodynamics to non-Abelian local internal symmetry groups, is well 
understood, using the notions of sections and connections in a fiber bundle.
The transfer of this formalism to quantum theory is highly nontrivial and, in
my opinion, not yet adequately understood. If we use the approach via algebras
of observables sketched above then the incorporation of the additional structure 
due to (not directly observable) local internal symmetries is obscured by the 
singular nature of points and lines used in the classical case. To handle this
we need knowledge about the short distance behavior (ultraviolet limit) of the 
theory. A few tentative suggestions concerning the notion of a quantum connection 
are given in ref. \cite{ref7}. I consider the clear understanding of how local internal
symmetries  can be incorporated in a well defined mathematical structure as one of the
most  important immediate aims on which many subsequent developments may hinge.
It constitutes, of course, a hybrid theory since the global nature of the geometric 
symmetries is kept. So it may not be of primary importance to clarify whether the 
continuum limit really exists.

\section{Retrospective and Perspectives}
Comparing the picture sketched so far with the discussions on the
interpretation of quantum theory seventy years ago we may note:

1. The `language of classical physics' stressed so much by Bohr as indispensable 
for the observer (to enable him to tell what was done and learned)
remains an essential ingredient but, if we disregard questions of convenience,
it may be reduced to the description of geometric relations in the placement
of various macroscopic bodies and the coarse events observed in space and time.
All further information is contained in the mathematical structure. The correspondence
principle needed to map the description into the realm of mathematical  symbols is
provided by the reference to classical space-time and its geometric  symmetry on both
sides. It is the correspondence (\ref{eq1}) together with the action of the translation
group, needed to characterize the passive nature of detectors by (\ref{eq4}). Apart from
this the global symmetry of space-time is needed in two respects. In a single experiment
because it studies the statistical relations  in an ensemble of many individual event
patterns which occur at different times and in the communication with other observers
who would like to test the results in a different region of space-time.

2. The indivisibility of a process emphasized by Bohr leads to the concept
of an event as an irreducible unit and it manifests itself also in the holistic aspect 
of the causal relations between events. The isolation of an individual process
as a distinguishable, coherent part in the history of the universe, a pattern of 
events which can be considered by itself without mentioning its ties with other 
parts, depends to some extent on the choice of the observer of how much
he wants to consider but this choice is limited by the requirement that it must lead
to a well defined conspicuous 
probability assignment for the total process, a requirement
which can be precisely fulfilled only in a steady environment.

3. The distinction between possibilities and facts which appears to be unavoidable 
in a formulation of indeterministic laws implies a distinction between future
and past. Bohr mentions the `essential irreversibility inherent in the very concept
of observation'. If the term observation does not mean that the ultimate responsibility 
for deciding what constitutes a fact is delegated to the consciousness of an
individual human being\footnote{I think this would be too unreliable to be useful for
the purposes of physics.} then we must accept the essential irreversibility inherent
in the concept of an event. This endows the `arrow of time' with an intrinsic 
significance in the physical theory and corresponds to a picture of reality as 
evolving in successive steps of a process with a moving boundary, separating 
past facts from future possibilities. It corresponds to the picture drawn
by the philosopher A.N. Whitehead \cite{ref8}. This does not conflict with the
existence of a time reversal symmetry of the theory which describes a symmetry
in the probability assignments for processes. The significance of the arrow
of time is encoded in the existing theory by the spectrum condition for the 
energy-momentum of states entering in the characterization of detectors
(which provide one contribution for the probability of an event). The time reversal 
operator, being anti-unitary, does not change the sign of the energy.

Turning now to perspectives for future development of the theory, we
might take a few hints from the preceding discussion. First, that all
symmetries should be considered as local but that we should not
associate the meaning of ´local´ with a point in a space--time continuum but
with a possible event. A pattern of events with a web of causal ties 
between them bears some analogy to a section in a fiber bundle whose
base space is the set of events and the typical fiber is a direct sum of
representations of the symmetry group. The causal ties provide the
connection. The dynamical law must then describe the probability
assignment for different possibilities of growth of such a pattern in
the evolution process in which possibilities turn into facts and the
boundary between past and future changes. Included in this task is the
determination of the subset of representations in the fiber of an
event, the generalized eigenvalue problem yielding  the relation
between masses, spins and charge quantum numbers. I shall not try to
elaborate on the many questions connected with such a picture and its
relation to existing formalism. This is beyond the scope of this paper
and the capabilities of its author.


\begin{thebibliography}{99}

\bibitem{ref1}  Ph.\ Blanchard and A. Jadczyk (eds.) \emph{Quantum Future},
Proc.\ Przesieka Conf.\ 1997, Springer Verlag, Heidelberg, 1999.

\bibitem{ref2} R. Bonifacio (ed.) \emph{Mysteries, Puzzles and Paradoxes in Quantum
Mechanics}, Proc.\ Lake Garda Conf.\ 1998, AIP Conf.\ Proc.\ 461.

\bibitem{ref3} Proc.\ Lake Garda Conf.\ 1999, to appear.

\bibitem{ref4} R. Omn\`es, \emph{The Interpretation of Quantum Mechanics},
Princeton University Press, 1994.

\bibitem{ref5} J.F. Clauser and M.A. Horne, {Phys. Rev. D 10}{526}{1974}.

\bibitem{ref6} R. Haag, \emph{Local Quantum Physics}, second edition.
Springer Verlag, Heidelberg, 1996.

\bibitem{ref7} D. Buchholz and R. Haag, \emph{The quest for understanding in
relativistic quantum physics}, preprint hep-th/9910243, to appear in 
J. Math.\ Phys., special issue. 

\bibitem{ref8} A.N. Whitehead, \emph{Process and Reality}, Macmillan Publishing Co.,
1927.

\end{thebibliography}
\end{document}